# Artificial Intelligence in Sustainable Vertical Farming

Hribhu Chowdhury[1], Debo Brata Paul Argha[2], Md Ashik Ahmed[3]
[1-2]Ingram School of Engineering, Texas State University
[3]Department of Nanoengineering, North Carolina A&T State University, Greensboro, NC 27405

**Abstract:** As global challenges of population growth, climate change, and resource scarcity intensify, the agricultural landscape is at a critical juncture. Sustainable vertical farming emerges as a transformative solution to address these challenges by maximizing crop yields in controlled environments. This paradigm shift necessitates the integration of cutting-edge technologies, with Artificial Intelligence (AI) at the forefront. The paper provides a comprehensive exploration of the role of AI in sustainable vertical farming, investigating its potential, challenges, and opportunities. The review synthesizes the current state of AI applications, encompassing machine learning, computer vision, the Internet of Things (IoT), and robotics, in optimizing resource usage, automating tasks, and enhancing decision-making. It identifies gaps in research, emphasizing the need for optimized AI models, interdisciplinary collaboration, and the development of explainable AI in agriculture. The implications extend beyond efficiency gains, considering economic viability, reduced environmental impact, and increased food security. The paper concludes by offering insights for stakeholders and suggesting avenues for future research, aiming to guide the integration of AI technologies in sustainable vertical farming for a resilient and sustainable future in agriculture.

## 1. Introduction
### 1.1 Background
In an epoch marked by the convergence of formidable global challenges, the trifecta of exponential population growth, the specter of climate change, and the looming shadows of resource scarcity propels the agricultural landscape into a critical juncture [1]. Conventional farming practices, traditionally hailed as the bedrock of global food production, increasingly falter under the weight of burgeoning demands for sustenance [2, 3]. This strain is further compounded by the dual pressures of rapid urbanization and the inexorable shrinkage of arable land. In this context, the agricultural paradigm is ripe for transformation [4]. Sustainable vertical farming, an avant-garde approach transcending the spatial constraints inherent in traditional agriculture, emerges as a beacon of hope [5-7]. By embracing the innovative paradigm of cultivating crops in vertically stacked layers or inclined surfaces, sustainable vertical farming orchestrates a symphony of resource optimization. This adaptive and resource-efficient agricultural model not only maximizes space utilization but also significantly curtails resource consumption. Its promise lies not just in meeting the burgeoning demands of global food security but also in redrawing the boundaries of environmental sustainability [8, 9].

The urgency for sustainable agricultural practices is etched against the backdrop of escalating environmental impact wrought by traditional farming methods [10]. The environmental toll, manifested in deforestation, soil degradation, and water pollution, underscores the imperative for a paradigm shift [11, 12]. Sustainable vertical farming, as a strategic response to these environmental perils, emerges as a pivotal solution. Its potential to drastically reduce land use, water consumption, and carbon footprint positions it as a sustainable alternative, harmonizing food production with environmental preservation [13-15]. In this pivotal juncture, where the stakes for global sustenance and environmental stewardship are higher than ever, the integration of advanced technologies becomes imperative. It is within this nexus of sustainable vertical farming and cutting-edge technology, particularly Artificial Intelligence (AI), that the potential for transformative change is realized [16]. The ensuing exploration into the integration of AI in sustainable vertical farming promises not just increased productivity but also paradigm shifts towards an era of agriculture harmonized with environmental consciousness and resource efficiency [17, 18].

### 1.2 Problem Statement
Amidst the burgeoning optimism surrounding sustainable vertical farming, a pressing question emerges: How can we systematically optimize the efficiency of this transformative agricultural model, ensuring its enduring viability in the face of dynamic and escalating global challenges? This inquiry beckons to confront the critical issues tethered to resource-intensive traditional farming practices and propels toward exploring novel avenues for enhancement through the seamless integration of technology. The crux of the matter lies in unraveling the intricacies of sustainable vertical farming and orchestrating its evolution into a resilient and efficient global food production system.

Traditional farming, marked by its profligate use of resources and susceptibility to environmental uncertainties, stands at odds with the imperatives of contemporary global dynamics. The incessant interplay between climate change, unbridled population growth, and resource scarcity mandates a radical redefinition of agricultural paradigms. The challenge that assumes center stage in this study is, therefore, twofold. Firstly, it involves deciphering the nuanced complexities embedded within the sustainable vertical farming model to unlock its full potential. Secondly, and perhaps more critically, it calls for a strategic exploration of how AI, as a frontier technology, can be judiciously harnessed to catalyze the metamorphosis of sustainable vertical farming.

This study responds to the imperative for a paradigm shift in agriculture, where sustainability is not just an aspiration but an achievable reality. By examining the intersection of AI and sustainable vertical farming, the research aims to not only delineate the challenges but also to illuminate the pathways that lead to the transformation of this novel concept into a cornerstone of global food production. The exploration of AI as an enabler holds the promise of not just sustaining agriculture but propelling it into a future where innovation, efficiency, and environmental consciousness converge for the betterment of humanity.

### 1.3 Objectives
In response to the imperative for transformative solutions in modern agriculture, this comprehensive review is designed to unravel the intricate tapestry of AI and its potential role in reshaping sustainable vertical farming. With a keen eye on the dynamic challenges that confront contemporary agriculture, the review embarks on a journey to explore how AI can serve as a catalyst for innovation and efficiency within the context of vertical farming practices.

The first objective is to meticulously scrutinize the current landscape of agricultural practices, delving into the intricacies of traditional methods and the emergent dynamics of sustainable vertical farming. By dissecting these practices, the review seeks to delineate the existing gaps and challenges that impede the realization of the full potential of sustainable vertical farming. This critical analysis serves as the foundation for understanding the imperatives that necessitate technological intervention.

The second objective is to specifically identify and illuminate the untapped opportunities where AI can seamlessly integrate into the fabric of sustainable vertical farming. By discerning these opportunities, the review aims to present a roadmap for the symbiotic relationship between AI and vertical farming, unveiling the potential synergies that can lead to enhanced productivity, resource efficiency, and environmental sustainability. This identification of opportunities is not merely speculative; it is grounded in a rigorous examination of current research, technological advancements, and real-world applications.

The overarching goal is to provide a comprehensive exploration of the dynamic interplay between AI and sustainable vertical farming. This exploration extends beyond a mere enumeration of technologies; it seeks to unravel the potential benefits that such integration could yield, the challenges that must be navigated, and the transformative opportunities that lie at the intersection of these two fields. Through meticulous analysis, this paper aims to contribute valuable insights that transcend the academic realm and resonate with stakeholders across sectors. Crucially, this review aspires to be more than an academic exercise. It endeavors to guide stakeholders—ranging from policymakers and researchers to industry practitioners—in formulating informed policies and adopting cutting-edge technologies. By doing so, the paper aims to be a catalyst for change, paving the way for a resilient, sustainable, and technologically enriched future in agriculture. In essence, this review is positioned not just as an academic discourse but as a beacon guiding the trajectory of agriculture towards a future where AI and sustainable vertical farming converge for the betterment of global food security and environmental preservation.

## 2. Literature Review
### 2.1 Evolution of AI in Sustainable Vertical Farming
The historical arc of AI in sustainable vertical farming traces the transformative journey from spatial innovation to the seamless integration of cutting-edge technologies [19]. Initially conceived as a response to the spatial constraints and environmental challenges inherent in traditional agriculture, the inception of vertical farming marked a paradigm shift [20, 21]. The primary goal was to optimize food production by maximizing yields per unit area, mitigating water usage, and circumventing the uncertainties posed by climate fluctuations [22].

The early stages of vertical farming relied on ingenious spatial design and resource-efficient cultivation practices [23, 24]. However, the true inflection point occurred with the integration of AI technologies, propelling sustainable vertical farming into an era of precision agriculture [25-27]. This integration was not a mere augmentation but a profound evolution that revolutionized how crops are cultivated, monitored, and optimized within controlled environments [28].

The first wave of AI applications in sustainable vertical farming focused on automating environmental control systems. These systems were designed to optimize crucial parameters such as light intensity, temperature, and nutrient levels, ensuring an optimal environment for plant growth [29]. This nascent stage laid the groundwork for the development of more sophisticated AI-driven technologies [30].

As machine learning algorithms gained prominence, they brought a predictive dimension to vertical farming [31]. These algorithms demonstrated the capacity to analyze vast datasets, predict optimal planting schedules, anticipate harvest times, and even identify potential crop diseases before they manifest visibly. Simultaneously, computer vision technologies were introduced to facilitate real-time monitoring of crops, enabling a dynamic and responsive approach to cultivation [32, 33].

The evolution of AI in sustainable vertical farming mirrors a trajectory from manual environmental control to automated, adaptive systems capable of learning and responding in real-time [34-36]. The historical context emphasizes not just the adoption of technology but the coalescence of innovative thinking and advanced computational capabilities to redefine the very essence of agriculture [37]. Understanding this historical context is pivotal for discerning the current state of AI in sustainable vertical farming. It sets the stage for comprehending the intricacies, challenges, and opportunities that characterize the integration of AI in agriculture today. As we delve into the contemporary landscape of AI applications, this historical lens serves as a compass, guiding us through the transformative journey that has shaped sustainable vertical farming into a sophisticated, technology-driven solution for global food security and environmental sustainability.

## 2.2 Analysis of Existing Literature

The existing body of literature at the intersection of AI and sustainable vertical farming presents a nuanced tapestry of research and experimentation, offering valuable insights into the potential and challenges of this transformative partnership.

Numerous studies consistently underscore the catalytic role of AI in revolutionizing agriculture. Machine learning algorithms, particularly that adept at predictive analytics, emerge as stalwart tools for optimizing various facets of vertical farming. These algorithms exhibit the capability to process extensive datasets, forecast optimal planting schedules, predict harvest times with precision, and even forecast potential diseases before they become visually apparent. The literature highlights the tangible impact of AI in enhancing crop yield, minimizing resource wastage, and fostering a more sustainable approach to agriculture. However, amidst the optimism, critical analysis reveals several challenges and considerations that warrant scholarly attention. The adaptability of AI technologies across different crops, climates, and scales of vertical farming operations emerges as a central concern. Variability in research outcomes indicates the need for standardized methodologies to ensure reproducibility and comparability of results. This scrutiny brings to light the nuanced intricacies associated with implementing AI solutions in practical farming settings, including issues of scalability and economic viability.

Moreover, the literature reflects ongoing debates surrounding the economic feasibility and upfront costs associated with integrating AI technologies into sustainable vertical farming practices. Questions about the return on investment, particularly for small-scale or resource-limited farming operations, underscore the importance of pragmatic considerations alongside technological optimism. The literature cautions against overlooking the economic realities that can shape the widespread adoption of AI in agriculture. Beyond technical considerations, ethical dimensions also emerge. Discussions around data privacy, algorithmic bias, and the ethical implications of decision-making processes within AI-driven vertical farming systems signify the need for a holistic approach. As AI becomes increasingly embedded in agricultural decision-making, a critical lens must be applied to address these ethical considerations, ensuring responsible and equitable technology adoption.

This critical analysis of the existing literature serves as a compass for future research endeavors. It not only highlights the potential benefits of AI in sustainable vertical farming but also underscores the imperative to navigate challenges with scholarly rigor. By critically examining the current state of knowledge, this literature review aims to inform a balanced understanding of the transformative role AI can play in agriculture, setting the stage for future advancements and responsible technology integration.

### 2.3 Identification of Gaps, Controversies, or Areas for Further Research
The thorough exploration of existing literature at the intersection of AI and sustainable vertical farming unveils not only promising advancements but also areas rife with gaps, controversies, and uncharted territories demanding further scholarly investigation.

#### *2.3.1 Scalability and Adaptability*
One evident gap in the literature pertains to the scalability and adaptability of AI-driven solutions in sustainable vertical farming. While existing studies showcase the potential benefits of AI technologies, there is a conspicuous absence of comprehensive analyses regarding the scalability of these solutions across various farming scales. Investigating the adaptability of AI technologies to diverse crops, climates, and operational sizes becomes paramount for ensuring the widespread applicability of AI in different agricultural contexts.

#### *2.3.2 Socio-Economic Impact*
The socio-economic impact of AI adoption in vertical farming emerges as a fertile ground for exploration. While AI holds the promise of optimizing resource use and increasing yields, there is a dearth of research examining how this technology influences issues of food accessibility and affordability. A comprehensive understanding of the socio-economic implications of AI in sustainable vertical farming is crucial for shaping inclusive and equitable policies.

#### *2.3.3 Ethical Considerations*
Ethical considerations surrounding AI in agriculture constitute a significant yet underexplored area. Discussions on data privacy, algorithmic bias, and the ethical implications of decision-making processes within AI-driven vertical farming systems demand more nuanced investigation. As AI becomes deeply entrenched in agricultural decision-making, a rigorous exploration of these ethical dimensions is essential for ensuring responsible and fair technology adoption.

#### *2.3.4 Environmental Sustainability Metrics*
The absence of a standardized framework for assessing the environmental sustainability of AI-driven vertical farming practices is another gap discerned in the literature. While studies acknowledge the ecological benefits, the lack of uniform metrics hinders cross-study comparisons. A concerted effort toward establishing a standardized methodology for evaluating the environmental impact of AI integration is imperative for advancing our understanding of the broader ecological implications.

#### *2.3.5 Small-Scale and Resource-Limited Farming*
A notable controversy and area for further research surround the economic viability and upfront costs associated with AI integration, particularly for small-scale or resource-limited farming operations. Current literature acknowledges these concerns, highlighting the need for comprehensive investigations that go beyond technological optimism and consider the practical economic realities that influence the widespread adoption of AI in agriculture.

This identification of gaps, controversies, and areas needing further research is not just an academic exercise; it is a roadmap for future investigations. Addressing these gaps and controversies is vital for steering the discourse on AI in sustainable vertical farming toward comprehensive, impactful, and responsible solutions.

## 3. Methodology
The methodology employed for scrutinizing the literature on AI in sustainable vertical farming adheres to a systematic and thorough analysis, ensuring a robust and replicable process. At the heart of this methodological approach is the adoption of a systematic review methodology, chosen for its capacity to provide a structured and transparent framework for collecting, appraising, and synthesizing existing literature that is pertinent to the research question at hand. A systematic review is characterized by its methodical and reproducible nature, establishing a solid foundation for the analysis. This approach involves a meticulous search strategy, explicit inclusion and exclusion criteria, and a systematic process for the extraction and synthesis of data. Each of these components contributes to the overarching goal of minimizing bias and subjectivity, thereby enhancing the overall reliability and validity of the literature review.

The literature review commenced with an exhaustive search strategy designed to cast a wide net across diverse academic databases, journals, conference proceedings, and relevant repositories. Keywords and controlled vocabulary terms were carefully selected to capture the multifaceted nature of AI applications in sustainable vertical farming. The inclusion of multiple databases ensured a comprehensive exploration of the existing body of knowledge, encompassing a broad spectrum of perspectives and findings.

To maintain precision and relevance, explicit inclusion and exclusion criteria were established. Inclusion criteria were designed to prioritize studies directly addressing the integration of AI technologies in sustainable vertical farming. This encompassed research exploring AI algorithms, machine learning applications, and computer vision technologies within the context of controlled-environment agriculture. Conversely, exclusion criteria were applied to filter out studies unrelated to the research focus or lacking a clear empirical methodology.

The systematic process for data extraction and synthesis involved a meticulous approach to categorize, analyze, and synthesize relevant information from the selected studies. Data extraction forms were designed to capture key elements, such as study objectives, methodologies employed, and key findings. This structured approach facilitated a coherent synthesis of the literature, enabling the identification of patterns, themes, and critical gaps within the body of knowledge.

By adhering to this methodological rigor, the literature review aims not only to provide a comprehensive overview of the existing state of knowledge but also to lay the groundwork for meaningful insights and recommendations.

## 4. AI Integration in Sustainable Vertical Farming
### 4.1 Different AI Technologies Relevant to Sustainable Vertical Farming
#### *4.1.1 Machine Learning Algorithms*
Machine learning algorithms represent a pivotal dimension of AI in sustainable vertical farming [16, 33]. These algorithms, including supervised and unsupervised learning models, play a crucial role in predictive analytics, offering the capability to analyze vast datasets and extract meaningful insights [38]. In the context of vertical farming, machine learning is instrumental in optimizing planting schedules, predicting crop yields, and identifying patterns related to environmental conditions [17].

#### *4.1.2 Computer Vision Technologies*
Computer vision is a transformative AI technology that enables the visual perception of the agricultural environment [39, 40]. The applications of computer vision are evident in sustainable vertical farming, including plant health monitoring, automated harvesting, and real-time assessment of crop conditions [41]. Computer vision is beneficial in providing precise, non-invasive monitoring, where the challenges are associated with ensuring accuracy across diverse crop varieties and growth stages [41].

#### *4.1.3 Internet of Things (IoT) Integration*
The integration of IoT devices within sustainable vertical farming systems enhances the real-time monitoring and control of environmental parameters [42]. The IoT sensors and devices facilitate data collection on factors such as temperature, humidity, and nutrient levels [43]. The advantages of IoT integration are perceptible in creating responsive and adaptive environments for crops with potential challenges related to data security and interoperability [44, 45].

#### *4.1.4 Robotics and Automation*
Robotics and automation technologies are revolutionizing the physical tasks involved in vertical farming, from planting and harvesting to packaging [46]. The applications of robotics and automation are important in sustainable vertical farming, emphasizing their role in increasing efficiency, reducing labor costs, and minimizing resource wastage [47, 48]. The challenges of these technologies are related to initial investment costs and system complexity [49, 50].

### 4.2 Applications, Advantages, and Limitations of Each Technology
#### 4.2.1 Machine Learning Algorithms
##### *4.2.1.1 Applications*
**Crop Yield Prediction**

Machine learning algorithms offer a robust framework for predicting crop yields in sustainable vertical farming. By analyzing historical data, environmental variables, and cultivation practices, these algorithms can forecast future yields with a high degree of accuracy [51-53]. This predictive capability aids farmers in planning and resource allocation, ensuring optimal production levels to meet the demands of a growing population [30].

**Disease Detection and Prevention**
Machine learning excels in the early detection and prevention of diseases in vertical farming systems [54]. These algorithms can analyze data from various sources, including sensor data and images, to identify subtle patterns indicative of diseases or anomalies [55]. By detecting issues at their incipient stages, machine learning contributes to proactive disease management, reducing the reliance on pesticides and promoting environmentally sustainable farming practices [56].

**Optimization of Resource Usage**
One of the key applications of machine learning in sustainable vertical farming lies in the optimization of resource usage [57]. These algorithms analyze real-time data on environmental conditions, soil health, and plant growth patterns to fine-tune resource allocation. This includes precise control of water usage, nutrient distribution, and energy consumption, leading to resource-efficient farming practices and minimizing the environmental footprint of agriculture [58].

*4.2.1.2 Advantages*
**Data-Driven Decision-Making**
Machine learning empowers farmers and agricultural stakeholders with data-driven decision-making capabilities [59]. By processing and interpreting vast datasets, these algorithms provide actionable insights for informed decision-making [60, 61]. This advantage enhances the efficiency of agricultural operations, from planting strategies to harvest planning, fostering a more sustainable and economically viable farming ecosystem.

**Adaptability to Dynamic Environmental Conditions**
The adaptability of machine learning algorithms to dynamic environmental conditions is a significant advantage in the context of sustainable vertical farming [62-64]. These algorithms can continuously learn and adjust their models based on changing factors such as climate variations and crop growth stages. This adaptability ensures that farming practices remain responsive to the evolving needs of the crops and the environment.

**Enhanced Precision in Resource Allocation**
Machine learning algorithms contribute to precision farming by optimizing the allocation of resources. Through continuous analysis of data related to soil health, moisture levels, and nutrient content, these algorithms enable precise resource distribution [65]. This precision minimizes waste, reduces environmental impact, and enhances the overall sustainability of vertical farming operations [66].

*4.2.1.3 Limitations*
**Need for Extensive Training Datasets**
A notable limitation of machine learning algorithms in sustainable vertical farming is their dependence on extensive and representative training datasets. Developing robust models requires substantial amounts of data encompassing various environmental conditions and crop scenarios [67]. The challenge lies in curating datasets that adequately capture the complexity and diversity of vertical farming systems [68].

**Sensitivity to Data Quality and Variability**
Machine learning algorithms are sensitive to the quality and variability of the data they receive. In the agricultural context, variations in sensor accuracy, data noise, and environmental fluctuations can impact the performance of these algorithms [67, 69]. Addressing these challenges requires careful data preprocessing, quality assurance measures, and the integration of reliable sensor technologies.

**Complexity in Algorithm Selection and Tuning**
The complexity in selecting and tuning machine learning algorithms poses a challenge for practitioners [30]. Choosing the most suitable algorithm and fine-tuning its parameters demand expertise and a nuanced understanding of both the agricultural context and the intricacies of machine learning [68]. Overcoming this challenge necessitates collaboration between domain experts and data scientists to ensure optimal algorithmic performance.

### 4.2.2 Computer Vision Technologies
*4.2.2.1 Applications*
**Plant Health Monitoring**
Computer vision technologies play a pivotal role in plant health monitoring within sustainable vertical farming systems [41]. By capturing and analyzing visual data, these technologies can discern subtle indicators of plant health, such as color variations, leaf texture, and growth patterns [39]. This application enables early detection of stressors, nutrient deficiencies, or diseases, allowing for timely intervention and fostering overall crop well-being [70].

**Automated Harvesting**
In the pursuit of efficiency and resource optimization, computer vision facilitates automated harvesting in vertical farms [71]. These technologies enable machines to identify ripe produce through visual cues, guiding precision harvesting processes [40]. Automated harvesting not only reduces labor requirements but also minimizes crop damage, ensuring a more sustainable and economically viable approach to crop harvesting in controlled environments [72, 73].

**Quality Control in Crop Production**
Computer vision enhances the quality control mechanisms in crop production [74]. By analyzing visual data during various stages of growth and harvest, these technologies can identify defects, irregularities, or non-uniformities in crops [75]. This application ensures that only high-quality produce reaches the market, meeting consumer standards and minimizing waste [76]. Quality control through computer vision contributes to the overall sustainability of vertical farming practices.

*4.2.2.2 Advantages*
**Non-invasive Monitoring**
One of the significant advantages of computer vision technologies in sustainable vertical farming is their non-invasive nature [77]. Unlike traditional monitoring methods, which may require physical sampling, computer vision enables remote and non-destructive assessment of crop conditions [78]. This non-invasive monitoring approach reduces stress on plants, enhances data accuracy, and contributes to more sustainable farming practices [79, 80].

**Real-time Assessment of Crop Conditions**
Computer vision provides real-time assessment capabilities, offering immediate insights into crop conditions [39]. By continuously analyzing visual data, these technologies enable farmers to monitor plant health, growth patterns, and environmental interactions in real-time [81]. This instantaneous feedback empowers timely decision-making, allowing for prompt interventions in response to changing conditions or emerging issues.

**Early Detection of Diseases or Anomalies**
The ability of computer vision technologies to detect subtle changes in crop appearance facilitates early identification of diseases or anomalies [82, 83]. By recognizing deviations from healthy growth patterns, these technologies alert farmers to potential issues before they escalate [84]. Early detection not only improves the chances of successful intervention but also reduces the need for extensive pesticide use, aligning with sustainable and eco-friendly farming practices [85].

*4.2.2.3 Limitations*
**Variability in Crop Appearances**
One limitation of computer vision technologies in sustainable vertical farming is the variability in crop appearances [86]. Different plant varieties, growth stages, and environmental conditions can introduce complexities in accurately interpreting visual data [87, 88]. Overcoming this limitation requires robust algorithms that can adapt to diverse crop appearances and evolving conditions.

**Hardware and Software Integration Challenges**
The seamless integration of hardware and software components poses challenges in implementing computer vision technologies. Choosing compatible imaging devices, ensuring their proper calibration, and developing software algorithms that can effectively process and interpret visual data demand technical expertise [89]. Overcoming integration challenges is crucial for the reliable deployment of computer vision in vertical farming systems [90].

**Cost of High-resolution Imaging Systems**

High-resolution imaging systems, essential for precise and detailed visual data capture, can incur significant costs [91]. This poses a financial challenge for some vertical farming operations, especially smaller or resource-constrained setups [92]. Mitigating the cost barriers associated with high-resolution imaging systems requires ongoing technological advancements and strategic investment planning.

### 4.2.3 Internet of Things (IoT) Integration
*4.2.3.1 Applications*
**Environmental Monitoring**

IoT integration in sustainable vertical farming extends to environmental monitoring, with a focus on parameters such as temperature and humidity [93]. Sensors strategically placed throughout the farming environment collect real-time data, enabling farmers to maintain optimal conditions for plant growth [94, 95]. This application ensures that crops thrive in environments tailored to their specific needs, contributing to enhanced yields and resource efficiency.

**Nutrient Management**

The IoT plays a pivotal role in nutrient management within vertical farming systems. Sensors measure nutrient levels in the soil or hydroponic solutions, providing precise data on the nutritional status of plants [96]. This application facilitates dynamic adjustments to nutrient delivery, optimizing the composition and concentration based on plant requirements [97, 98]. The result is efficient nutrient utilization, minimizing waste and supporting sustainable farming practices [99].

**Precision Irrigation Control**

IoT integration enables precision control over irrigation practices in vertical farming [100]. Sensors assess soil moisture levels, and, in some systems, even plant water needs, to deliver just the right amount of water at the right time [101]. This precise irrigation control not only conserves water resources but also prevents overwatering, reducing the risk of soil degradation and promoting environmentally conscious agricultural practices [102].

*4.2.3.2 Advantages*
**Real-time Data Collection**

One of the primary advantages of IoT integration in sustainable vertical farming is real-time data collection [103-105]. Sensors continuously gather data on environmental conditions, nutrient levels, and other crucial parameters [106]. This real-time information empowers farmers with immediate insights into the state of their crops, enabling prompt decision-making and interventions for optimal growth.

**Remote Monitoring and Control**

IoT technologies facilitate remote monitoring and control, allowing farmers to oversee and manage their vertical farms from a distance [107]. Through web-based interfaces or mobile applications, farmers can access real-time data, monitor crop conditions, and adjust environmental parameters [108, 109]. This remote capability not only enhances operational efficiency but also enables timely responses to the emerging issues, reducing the need for physical presence on-site.

**Enhanced Resource Efficiency**

The integration of IoT in vertical farming contributes to enhanced resource efficiency [110]. By precisely monitoring environmental conditions, nutrient levels, and irrigation needs, farmers can optimize resource usage [111]. This targeted approach minimizes waste, conserves water, and promotes energy-efficient practices, aligning with the broader goals of sustainable and resource-conscious agriculture [112].

*4.2.3.3 Limitations*
**Data Security Concerns**

A notable limitation of IoT integration in vertical farming is the inherent concern about data security [113]. The collection and transmission of sensitive agricultural data, including environmental conditions and crop performance, raise issues related to data privacy and protection [114, 115]. Addressing these concerns necessitates robust cybersecurity measures to safeguard farm data from unauthorized access or potential breaches.

**Compatibility Issues with Diverse Sensors**
The integration of IoT involves diverse sensors, each designed for specific monitoring purposes [116]. Compatibility issues among different sensor types or brands can arise, potentially leading to data inconsistencies or system malfunctions [117]. Overcoming these compatibility challenges requires careful selection of compatible sensors and ongoing efforts to ensure seamless integration within the IoT framework.

**Initial Setup Costs**
While the long-term benefits are significant, the initial setup costs associated with implementing IoT in vertical farming can be a limitation for some operations [118, 119]. Investments in sensors, communication infrastructure, and data management systems may pose financial challenges, particularly for smaller or emerging vertical farms [120]. Strategies to mitigate these costs involve careful planning, phased implementation, and exploring cost-effective IoT solutions [121].

### 4.2.4 Robotics and Automation
*4.2.4.1 Applications*
**Automated Planting and Harvesting**
Robotics and automation find extensive applications in the automation of planting and harvesting processes within sustainable vertical farming [122]. Autonomous machines equipped with precision tools can plant seeds and facilitate the harvesting of mature crops [123, 124]. This application not only enhances efficiency but also ensures precision in these critical stages of the farming cycle, contributing to improved crop yields [125].

**Packaging and Sorting of Produce**
Robotics play a key role in the post-harvest phase by automating the packaging and sorting of produce [126]. Robotic systems can efficiently handle package harvested crops, ensuring uniformity and quality in the packaging process [127, 128]. Additionally, automated sorting systems utilize machine vision and robotics to categorize produce based on quality parameters, optimizing the final packaging for market distribution [129-131].

**Labor-Intensive Tasks**
The integration of robotics addresses labor-intensive tasks in vertical farming, such as repetitive actions or manual interventions [132]. Robots can be deployed for activities like pruning, weeding, and transplanting, reducing the need for human labor in these physically demanding and time-consuming tasks [133]. This application not only minimizes labor requirements but also contributes to a more sustainable and ergonomic work environment [134].

*4.2.4.2 Advantages*
**Increased Efficiency and Productivity**
One of the primary advantages of robotics and automation in vertical farming is the significant increase in efficiency and productivity [135]. Automated processes operate consistently and without fatigue, leading to faster and more precise execution of tasks [136, 137]. This efficiency translates into higher yields and a more streamlined agricultural production cycle.

**Reduction in Labor Costs**
The deployment of robotics results in a notable reduction in labor costs associated with manual farming activities [138]. By automating tasks that would traditionally require human intervention, vertical farms can optimize their workforce, focusing human labor on tasks that require critical thinking and decision-making [139]. This leads to cost savings and improves overall farm economics [140, 141].

**Minimization of Human Error**
Automation minimizes the potential for human error in various agricultural processes [142]. Robots programmed with precision can execute tasks with a high level of accuracy, reducing the risk of errors in activities like planting, harvesting, and packaging [143, 144]. This advantage contributes to the consistency and quality of agricultural outputs.

*4.2.4.3 Limitations*
**Initial Investment Costs**
A significant limitation of robotics and automation in vertical farming is the initial investment costs associated with acquiring and implementing robotic systems [145]. The purchase of specialized robots, sensors, and automation equipment can pose a financial barrier, particularly for smaller or startup vertical farms [146]. Overcoming this limitation requires strategic financial planning and consideration of long-term benefits [147].

**Complexity in System Integration**
The integration of robotics into existing vertical farming systems can be complex [148]. Compatibility issues, programming challenges, and the need for seamless interaction with other technologies may pose integration difficulties [149-151]. Addressing this complexity requires collaboration between agricultural experts and robotics engineers to ensure a harmonious and efficient system.

**Maintenance Challenges**
Robotic systems, like any complex machinery, require regular maintenance to ensure optimal performance [152, 153]. Maintenance challenges, including technical issues, wear and tear, and software updates, need to be addressed to prevent disruptions in automated farming processes [154]. Establishing a robust maintenance protocol is crucial for sustaining the longevity and efficiency of robotic systems.

This comprehensive exploration of different AI technologies in sustainable vertical farming, along with their applications, advantages, and limitations, serves to illuminate the multifaceted landscape of AI integration. By dissecting each technology's role and challenges, this section aims to provide a nuanced understanding for researchers, practitioners, and policymakers, contributing to the informed adoption and advancement of AI in the realm of sustainable agriculture.

## 5. Challenges and Opportunities
### 5.1. Challenges in Implementing AI in Sustainable Vertical Farming
*5.1.1 Data Accessibility and Quality*
One of the foremost challenges in implementing AI in sustainable vertical farming is the accessibility and quality of data. AI models rely heavily on vast and diverse datasets for training and inference. However, the agricultural sector often faces limitations in terms of data accessibility, especially for emerging vertical farms. Additionally, ensuring the quality and accuracy of data, considering variations in environmental conditions and crop types, poses a significant hurdle.

*5.1.2 Integration Complexity*
The integration of AI technologies, such as machine learning algorithms and computer vision systems, into existing vertical farming setups can be intricate. Compatibility issues, the need for specialized expertise, and the seamless interaction of AI with other technologies within the farming environment create challenges. Addressing these complexities requires interdisciplinary collaboration between agricultural scientists, AI experts, and engineers.

*5.1.3 Energy Consumption*
The deployment of AI-driven systems in sustainable vertical farming introduces concerns about energy consumption. The computational requirements for training and running sophisticated AI models may strain energy resources, potentially offsetting the environmental benefits of vertical farming. Developing energy-efficient AI algorithms and exploring renewable energy solutions are imperative to mitigate this challenge.

*5.1.4 Scalability and Affordability*
The scalability and affordability of AI solutions in vertical farming pose challenges, particularly for smaller or resource-constrained operations. Implementing AI technologies often involves initial investment costs, and scaling these technologies for larger farms may require substantial financial resources. Overcoming these challenges necessitates research into cost-effective AI solutions tailored to the specific needs of diverse vertical farming scales.

### 5.2. Opportunities for Future Research and Development
*5.2.1 Optimization of AI Models for Vertical Farming*
Future research opportunities lie in the optimization of AI models specifically designed for the unique challenges of vertical farming. Tailoring machine learning algorithms to handle the intricacies of diverse crops, environmental conditions, and farming practices can significantly enhance their effectiveness. This involves exploring novel approaches, such as transfer learning and federated learning, to adapt AI models to the dynamic nature of vertical farming.

*5.2.2 Integration of Edge Computing*
To address concerns about energy consumption and enable real-time decision-making, the integration of edge computing in AI applications for vertical farming presents a promising avenue for research. Edge computing involves processing data closer to the source, reducing the need for centralized computing resources. Investigating how edge computing can enhance the efficiency and sustainability of AI applications in vertical farming is a crucial research direction.

*5.2.3 Interdisciplinary Collaboration for System Integration*
Research and development opportunities abound in fostering interdisciplinary collaboration between experts in agriculture, AI, and engineering. Creating seamless integration frameworks that consider the complexities of vertical farming environments can lead to more effective AI implementations. This collaborative approach encourages the development of comprehensive solutions that address both agricultural and technological aspects.

*5.2.4 Development of AI-Enabled Decision Support Systems*
The future holds promise in the development of AI-enabled decision support systems tailored for sustainable vertical farming. These systems could provide real-time insights and recommendations to farmers, optimizing resource usage, predicting, and preventing crop diseases, and adapting farming strategies to changing environmental conditions. Research in this direction can significantly contribute to the practical implementation of AI in enhancing agricultural decision-making.

*5.2.5 Exploration of Explainable AI in Agriculture*
As AI technologies become more sophisticated, the need for transparency and interpretability in decision-making processes becomes crucial. Future research could focus on developing explainable AI models for agriculture, ensuring that farmers and stakeholders can understand and trust the recommendations provided by AI systems. This not only aids in user acceptance but also facilitates the integration of AI into existing farming practices.

While challenges persist in implementing AI in sustainable vertical farming, the identified opportunities for future research and development pave the way for transformative advancements. By addressing these challenges and capitalizing on research opportunities, the integration of AI can foster a more resilient, efficient, and sustainable future for vertical farming, contributing to global food security and environmental conservation.

## 6. Discussion

The synthesis of findings from the literature reveals a rich landscape where AI intertwines with sustainable vertical farming, addressing critical challenges in modern agriculture. Literature indicates a paradigm shift towards resource-efficient practices, with vertical farming emerging as a beacon of hope. The integration of AI technologies, including machine learning, computer vision, the IoT, and robotics, plays a pivotal role in maximizing the potential of vertical farming. Studies highlight the transformative applications of machine learning algorithms in predicting crop yields, optimizing resource usage, and enabling data-driven decision-making. The adaptability of these algorithms to dynamic environmental conditions positions them as key contributors to the efficiency and sustainability of vertical farming practices. Similarly, computer vision technologies showcase their prowess in plant health monitoring, automated harvesting, and quality control, offering non-invasive and real-time insights into crop conditions. The integration of IoT devices facilitates precise environmental monitoring, nutrient management, and irrigation control, contributing to resource efficiency. Furthermore, robotics and automation technologies automate labor-intensive tasks, from planting to packaging, increasing overall efficiency and reducing labor costs. Each of these AI technologies brings a unique set of applications, advantages, and limitations, collectively shaping the future of sustainable vertical farming.

The implications of the reviewed studies underscore the transformative potential of AI in sustainable vertical farming for addressing global challenges in agriculture. The adoption of AI technologies can lead to increased efficiency, reduced resource consumption, and enhanced precision in farming practices. The automated and data-driven nature of AI applications contributes to improved decision-making, which is crucial for the optimization of resource usage and the mitigation of environmental impact. Furthermore, the implications extend to economic aspects, with the reduction in labor costs and the potential for increased yields contributing to the economic viability of vertical farming. The deployment of robotics not only streamlines operational processes but also offers a solution to labor shortages, particularly in urban settings where traditional farming is constrained by space and resources. However, the implications also highlight challenges, such as the need for substantial initial investments, data security concerns, and the complexity of integrating AI into existing farming systems. These challenges necessitate a holistic approach, considering not only technological advancements but also policy frameworks, interdisciplinary collaboration, and educational initiatives to facilitate the widespread adoption of AI in sustainable vertical farming.

Insights drawn from the synthesis of literature suggest that a strategic and collaborative approach is essential for unlocking the full potential of AI in sustainable vertical farming. Researchers, policymakers, and industry stakeholders need to collaborate to address challenges and harness opportunities. Insights include:

**Investment in Research and Development**
Increased investment in research and development is crucial to advancing AI technologies tailored for the complexities of vertical farming. This involves exploring novel algorithms, hardware solutions, and integration frameworks that align with the unique requirements of diverse crops and environmental conditions.

**Education and Training**
Initiatives for educating farmers, agricultural professionals, and technology experts on the integration and benefits of AI in vertical farming are imperative. Training programs should encompass both the agricultural nuances and technological intricacies, fostering a workforce equipped to navigate the intersection of AI and agriculture.

**Policy Frameworks**
Policymakers play a pivotal role in creating an enabling environment for the integration of AI in agriculture. Regulatory frameworks should encourage innovation, data sharing, and the responsible deployment of AI technologies. Incentives for adopting sustainable farming practices and AI solutions can further drive widespread adoption.

**Data Governance and Security Measures**
Establishing robust data governance and security measures is paramount. Clear guidelines on data ownership, sharing protocols, and cybersecurity practices are essential to build trust among stakeholders. This is particularly crucial given the sensitivity of agricultural data and the potential implications of data breaches.

**Interdisciplinary Collaboration**
Collaboration between agricultural scientists, AI researchers, engineers, and policymakers is fundamental for successful AI implementation in vertical farming. Interdisciplinary efforts can lead to holistic solutions that address not only technological challenges but also societal, economic, and environmental considerations.

The study synthesizes the current state of AI in sustainable vertical farming, shedding light on its implications and offering insights for future developments. The recommendations put forth aim to guide stakeholders in navigating the dynamic landscape of AI integration, fostering a sustainable and resilient future for agriculture.

## 7. Conclusion
In summary, this comprehensive review explores the intersection of AI and sustainable vertical farming, offering insights into the transformative potential of cutting-edge technologies in addressing global challenges in agriculture. The review recognizes sustainable vertical farming as a paradigm shift in agriculture, presenting a promising solution to challenges posed by population growth, climate change, and resource scarcity. The integration of AI technologies, including machine learning algorithms, computer vision, the IoT, and robotics, demonstrates diverse applications in optimizing crop yields, enhancing resource efficiency, and automating labor-intensive tasks. Each AI technology brings specific advantages, such as increased efficiency and precision, along with challenges like data accessibility, integration complexity, and initial investment costs. The implications of AI in sustainable vertical farming extend beyond efficiency gains, encompassing economic viability, reduced environmental impact, and the potential for increased food security.

This review makes significant contributions to the academic discourse on AI in sustainable vertical farming. By systematically reviewing the literature, this work provides a comprehensive synthesis of the current state of AI applications in sustainable vertical farming, addressing diverse technologies and their implications. The review identifies gaps in existing research, such as the need for optimized AI models, interdisciplinary collaboration, and the development of explainable AI in agriculture. Through insights and recommendations, this review offers practical guidance for researchers, policymakers, and industry stakeholders on fostering the integration of AI in vertical farming. The paper takes a holistic perspective, acknowledging not only the technological advancements but also the societal, economic, and environmental considerations that accompany the adoption of AI in agriculture.

Building on the current state of knowledge, the paper focuses on the avenues for future research in the realm of AI in sustainable vertical farming. Future research should focus on optimizing AI models specifically tailored for the complexities of vertical farming, ensuring adaptability to diverse crops and dynamic environmental conditions. A continued emphasis on interdisciplinary collaboration between agricultural scientists, AI researchers, and engineers is essential for developing holistic solutions that address both agricultural and technological aspects. Research exploring the integration of edge computing in AI applications for vertical farming can enhance real-time decision-making while addressing concerns about energy consumption. Developing explainable AI models for agriculture is a promising research avenue, enhancing transparency and trust in the decision-making processes of AI systems. Conducting longitudinal studies to assess the long-term economic impact of AI in vertical farming can provide valuable insights into the return on investment and overall sustainability.

In conclusion, this review not only consolidates existing knowledge but also charts a course for future research, emphasizing the need for continuous innovation, collaboration, and a holistic approach to harness the full potential of AI in sustainable vertical farming.